\documentclass[twocolumn,amsmath,amssymb,prl]{revtex4}

\usepackage{graphicx}
\usepackage{comment}
\usepackage[normalem]{ulem}
\usepackage{color}
\usepackage[colorlinks=true,linkcolor=blue,anchorcolor=blue,citecolor=blue]{hyperref}%
\usepackage{amsmath}
\usepackage{soul}
\usepackage{gensymb}

\definecolor{darkpastelgreen}{rgb}{0.01, 0.75, 0.24}

\makeatletter
\def\blfootnote{\xdef\@thefnmark{}\@footnotetext}
\makeatother

\begin{document}
%
%
\title{Thermoelectric evidence of the electronic structure changes from the charge-density-wave transition in FeGe}

\author{Kaila Jenkins$^{1}$, Yuan Zhu$^{1}$, Dechen Zhang$^{1}$, Guoxin Zheng$^{1}$, Kuan-Wen Chen$^{1}$, Aaron Chan$^{1}$, Sijie Xu$^{2,3}$, Mason L. Klemm$^{2,3}$, Bin Gao$^{2,3}$, Ming Yi$^{2,3}$, Pengcheng Dai$^{2,3}$, }
\author{Lu Li$^{1}$}
\email{luli@umich.edu}

\affiliation{
    $^1$Department of Physics, University of Michigan, Ann Arbor, MI 48109, USA.\\
    $^2$Department of Physics and Astronomy, Rice University, Houston, TX 77005, USA.\\
    $^3$ Rice Laboratory for Emergent Magnetic Materials and Smalley-Curl Institute, Rice University, Houston, TX 77005, USA.
}

\date{\today}
\begin{abstract}
Kagome metals provide a material platform for probing new correlated quantum phenomena due to the naturally incorporated linear dispersions, flat bands, and Van Hove singularities in their electronic structures. Among these quantum phenomena is the charge density wave (CDW), or the distortion of the lattice structure due to the motion of correlated electrons through the material. CDWs lower the energy of the compound, creating an energy gap that facilitates behaviors akin to superconductivity, nonlinear transport, or other quantum correlated phenomena. The kagome metal FeGe has been shown to host a CDW transition at approximately 100 K, and its occurrence is strongly influenced by the sample annealing conditions. However, a notable gap in the literature is the lack of clear thermoelectric transport evidence for electronic structure changes associated with this CDW transition. Here we present evidence of electron behavior modification due to annealing disorder via thermoelectric measurements on FeGe crystals presenting a CDW transition and those without a CDW. The observed Nernst effect and Seebeck effect under sufficient annealing demonstrate modified electrical transport properties resulting from induced disorder, including a change in carrier sign and an enhancement of the Nernst effect due to the CDW. Our results provide evidence of multiple phase transitions, which confirms the influence of CDW on the thermal properties of FeGe and demonstrates the suppression of CDW with sufficient disordering.
\end{abstract}


\maketitle                   
\renewcommand{\thesubsection}{\Alph{subsection}}
\renewcommand{\thesubsubsection}

 Novel electronic and magnetic materials have garnered considerable interest in recent years because of their environmental, technological, and practical benefits. In particular, novel thermoelectric materials are materials that exhibit enhanced and attractive thermoelectric behaviors \cite{thermalbook} such as higher electrical conductivity, lower thermal conductivity, or generally unique thermoelectric signatures such as the anomalous Nernst effect found in strongly spin-orbit-coupled materials \cite{Nagaosa2010, Onada2008, ikhlas2017, yang2020, guin2019, Madhogaria2023, Zhang2021}. The Nernst effect \cite{Behnia, Behnia2016, Sondheimer1948} provides evidence of mobile vortices in superconductors \cite{zhu2010, Wang2006, Onose2007, Chen2020}. Fundamentally, the quasiparticle thermopower (Seebeck effect) provides a different test of the carrier types \cite{Ziman2001}, and the quasiparticle Nernst effects are observed in semimetals with small Fermi surfaces \cite{Behnia2015}. The thermoelectric effect thus provides unique insight into the underlying physical properties of kagome materials \cite{Jiang2021, Xu2023} and other strongly correlated electron materials \cite{Sun2013}.

\begin{figure}[!htb]
	\begin{center}
		\includegraphics[width= \columnwidth ]{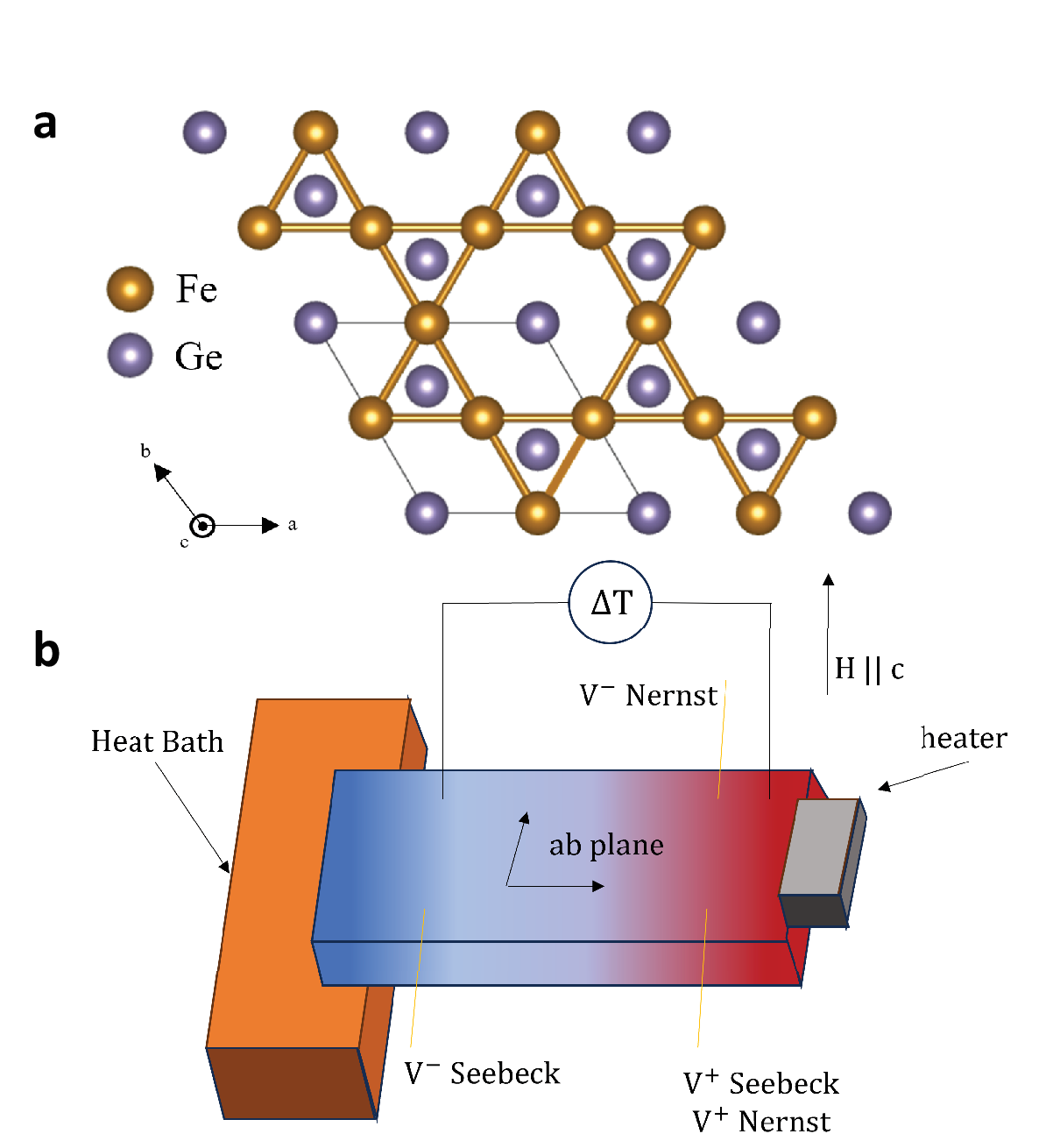}
		\caption{(a) Crystal structure of Kagome magnet FeGe. (b) Experimental configuration for FeGe thermoelectric measurement. Gold leads indicate electrical contacts, with the sign convention labeled. The sample is oriented with the magnetic field applied along the crystal $c$-axis. }
  
		\label{Fig1}
	\end{center}
\end{figure}

The kagome material FeGe (the B35 phase) was reported to host a CDW transition around 100 K in the antiferromagnetic phase \cite{Teng2022, Yin2022, Teng2023, Shao2023}. Evidence for this transition has been observed with scanning tunneling microscopy \cite{Yin2022},  photoemission spectroscopy \cite{Teng2023,Oh2025,Zhao2025}, magnetic susceptibility, heat capacity, and the electrial transport measurements \cite{Teng2022}. There is also a strong spin-lattice-charge coupling across the CDW transition \cite{PhysRevLett.133.04650}.
The amorphous and thin film of the cubic (the B20 phase)  FeGe samples have been reported to exhibit the anomalous Hall effect \cite{Bouma2020, Porter2024}. 
However, the anomalous Hall effect in kagome lattice FeGe was only found at a temperature well below the CDW transition, where incommensurate magnetic order was established below $T_{cant}$ ($T_{cant}<T_{CDW}$)  \cite{Bernhard84, Lebing24, Wu2024, Ziyuan24, Shi2024, Klemm2024}.
In addition, the suppressibility of CDW in kagome FeGe has been investigated, specifically with regard to annealing-induced CDW suppression and its effects on magnetic ordering \cite{Wu2024,  Ziyuan24, Shi2024, Klemm2024}.

However, despite the evidence for unique thermoelectic behavior in other topological and kagome compounds \cite{Zhu2024, Asaba2021, Roychowdhury2022, Bhandari2024, Zeng2022, Ceccardi2023, Sharma2018}, Weyl materials \cite{Sharma2017, Watzman2018, Xu2020ratio}, and evidence of anomalous Nernst effects and novel thermoelectric behavior in various ferrous materials \cite{lee2004, Miyasato2007, Pu2008, Chuang2017, Ramos2014}, no other studies have published observations regarding the unique thermal features linked to the charge and magnetic ordering present in this compound. In particular, a question remains as to whether there are any sharp features in transport properties associated with the transitions. This letter seeks to fill this gap. Additionally, this work seeks to identify the role of disorder in both thermal behavior and magnetic ordering, as disorder has been shown to influence Nernst effects \cite{Ding2019}. We report the finding of a strong Nernst effect in kagome FeGe, providing evidence of its features for the proposed CDW. We demonstrate here that kagome FeGe exhibits evidence of a charge density wave in thermopower measurements and a strong Nernst effect, and that this behavior is strongly suppressed when CDW order is removed through the process of high temperature annealing \cite{Wu2024,  Ziyuan24, Shi2024, Klemm2024}. We also provide evidence here for magnetic disordering when the CDW feature is suppressed. 

Single crystals were grown using Sn flux, and then annealed in separate batches under different temperature conditions. One batch was annealed at 320 $\degree$C for 96 hours, and another batch was annealed at 560 $\degree$C for 96 hours \cite{Klemm2024}. The orientation of the annealed samples was confirmed using X-ray diffraction. Samples were cut and polished to approximately 1 mm x 0.5 mm x 0.1 mm, perpendicular to the $c$-axis, and each was mounted on a copper block as a heat bath. Thermocouple type-E wires and gold-wire electrical contacts were mounted along the $a$-$b$ plane. The negative Constantan leads were electrically shorted together. A 1 k$\Omega$ resistor was mounted on top of the sample to provide the necessary heat gradient for the measurement, and a layer of thermal epoxy was applied between the sample and the heater to prevent electrical contact. Each lead was mounted to its own nylon pillar using silver epoxy, and the electrical and thermal contacts were made to the PPMS puck using silver paint. Thermal epoxy was applied to each of the thermocouple joints, which were adhered to the sample without making electrical contact. This experimental setup is illustrated in Fig. \ref{Fig1}, along with the FeGe crystal structure. 

Measurements were taken using a Quantum Design Physical Property Measurement System (PPMS) Dynacool with a maximum field of 14 T. The magnetic field is oriented normal to the crystal plane surface. In FeGe, a square wave current excitation with an amplitude of 1.6 mA and a frequency of 0.04 Hz was utilized to energize the heater. The Nernst effect is measured perpendicular to the direction of the temperature gradient, and the Seebeck effect is measured along the direction of the temperature gradient. Each thermoelectric voltage signal was obtained by subtracting the bottom, heater-off signal, from the top, heater-on signal. The experiment for each sample consisted of several fixed-temperature field sweeps, varying the temperature in 5 K increments until after the anticipated 100 K charge density wave, and then with a slightly coarser temperature granulation after this feature. 

\begin{figure*}[!htb]
	\begin{center}
		\includegraphics[width = \textwidth ]{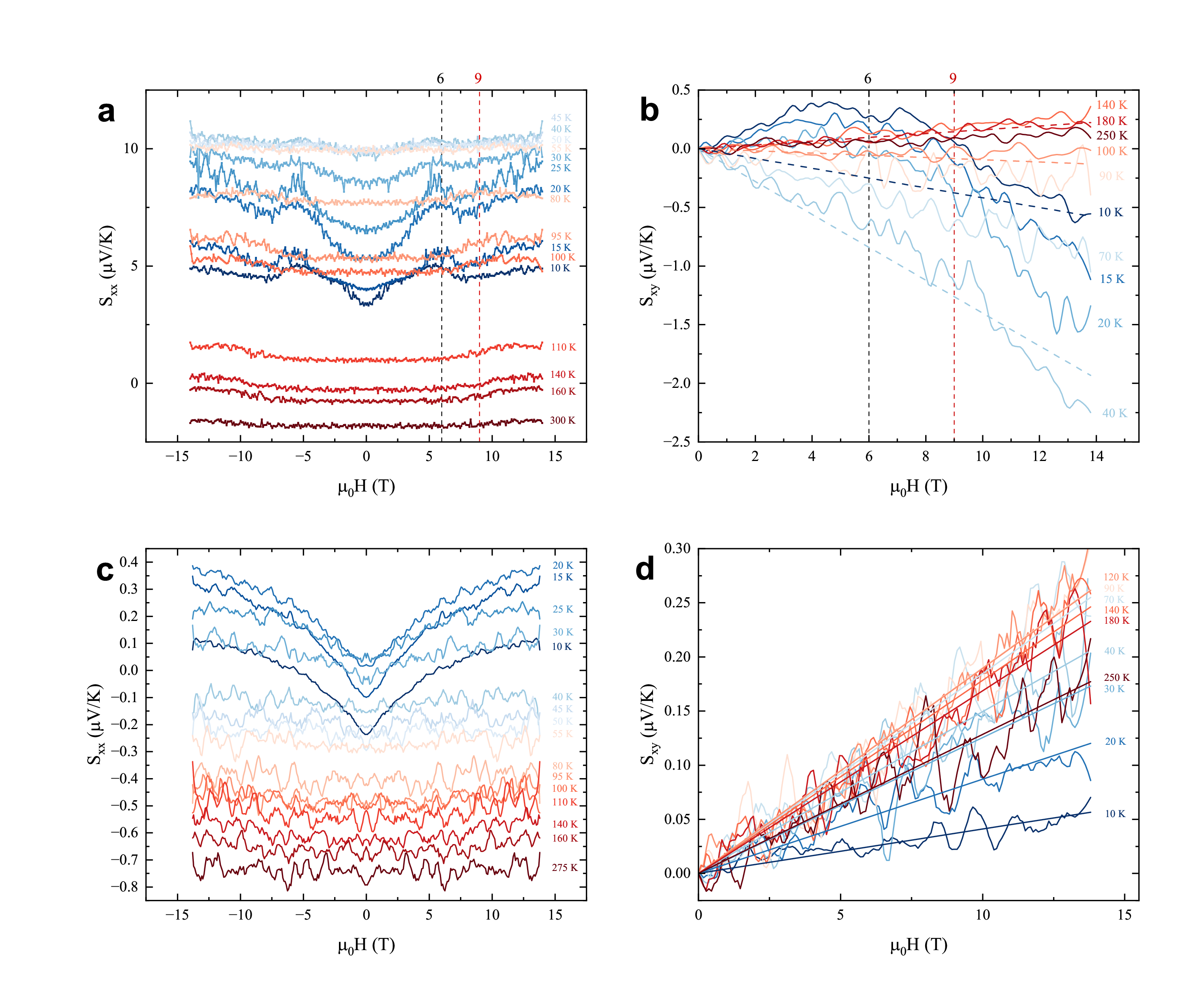}
		\caption{ (a, b) Thermopower signal $S_{xx}$ and the Nernst signal $S_{xy}$ of the 320 $\degree$C annealed sample as a function of the magnetic field at selected temperatures, along with selected linear fits as dotted lines. (c, d) Thermopower signal $S_{xx}$ and the Nernst signal $S_{xy}$ of the 560 $\degree$C annealed sample as a function of the magnetic field at selected temperatures, plotted alongside linear fits of the antisymmetrized transverse signal for clarity, as well as defining the Nernst coeffienct $\nu$.      
        } 
		\label{Fig2}
	\end{center}
\end{figure*}

The signal evolution of both the Seebeck signal, $S_{xx}$ and Nernst signal, $S_{xy}$, for both conditions as a function of temperature and magnetic field are shown in Fig. \ref{Fig2}. As can be noted in \ref {Fig2}a, the thermopower signal shows two distinctive features under magnetic fields, one occurring around 6 T and the other around 9 T. Comparing it with the early report, we identify these features as the spin-flop transitions \cite{Wu2024, Klemm2024}. These features appear to be present in the 320 $\degree$C annealed sample, but do not appear in the thermopower data for the 560 $\degree$C annealed sample, although the field-induced spin-flop transitions occur in both samples \cite{Klemm2024}. The small magnitude of both of these signals and their corresponding noise levels should be taken into consideration as a possible caveat. Also of note regarding the signal strength, the magnitude of the thermopower signal is an order of magnitude larger for the 320 $\degree$C sample for $T < T_{CDW}$, but is of a similar order of magnitude for $T > T_{CDW}$.

To understand the trend of the Nernst effect in this material and its dependence on annealing, fixed-temperature field sweeps were performed for several different temperatures. After deriving the raw Nernst signal $S_{xy}^0(H)$ from the raw signal, the signal was anti-symmetrized according to $\frac{S_{xy}^0(H) - S_{xy}^0(-H)}{2}$, with $H$ the applied magnetic field applied along the $c$-axis. This is done to isolate the true Nenrst component of the signal and to remove any artifacts or contributions from the longitudinal voltage. The resulting Nernst effect signal $S_{xy}$ for each of the annealing conditions is shown in Fig. \ref{Fig2}b and Fig. \ref{Fig2}d. Furthermore, the zero-field slope of $S_{xy}-\mu_0H$ is defined as the Nernst coefficient $\nu$, which will be discussed later.

The Nernst effect in both annealed samples has a generally linear trend, with some curvature appearing in the signal from the 320 $\degree$C annealed sample. Again, this curvature appears to correspond to the previously noted spin-flop features, and the slope has its largest magnitude around 40-45 K before gradually waning. This trend is notably absent in the 560 $\degree$C annealed sample signal, which does appear to have a maximum slope between 120 and 140 K, but then decreases in magnitude, and does not exhibit any other notable magnetic features. In addition to the general linear trend, the 320$\degree$C signal also exhibits a bump feature around 6 T, which is strongest between 40 and 45 K, and evolves with increasing temperature. This feature is likely a consequence of the 6 T spin-flop phase. However, as this signal is small, the noise is too prominent to see evidence of the 9 T spin-flop transition. The bump feature noted in the 320 $\degree$C sample does not appear to be present under the  560 $\degree$C  annealing condition.

\begin{figure}[!htb]
	\begin{center}
		\includegraphics[width= \columnwidth ]{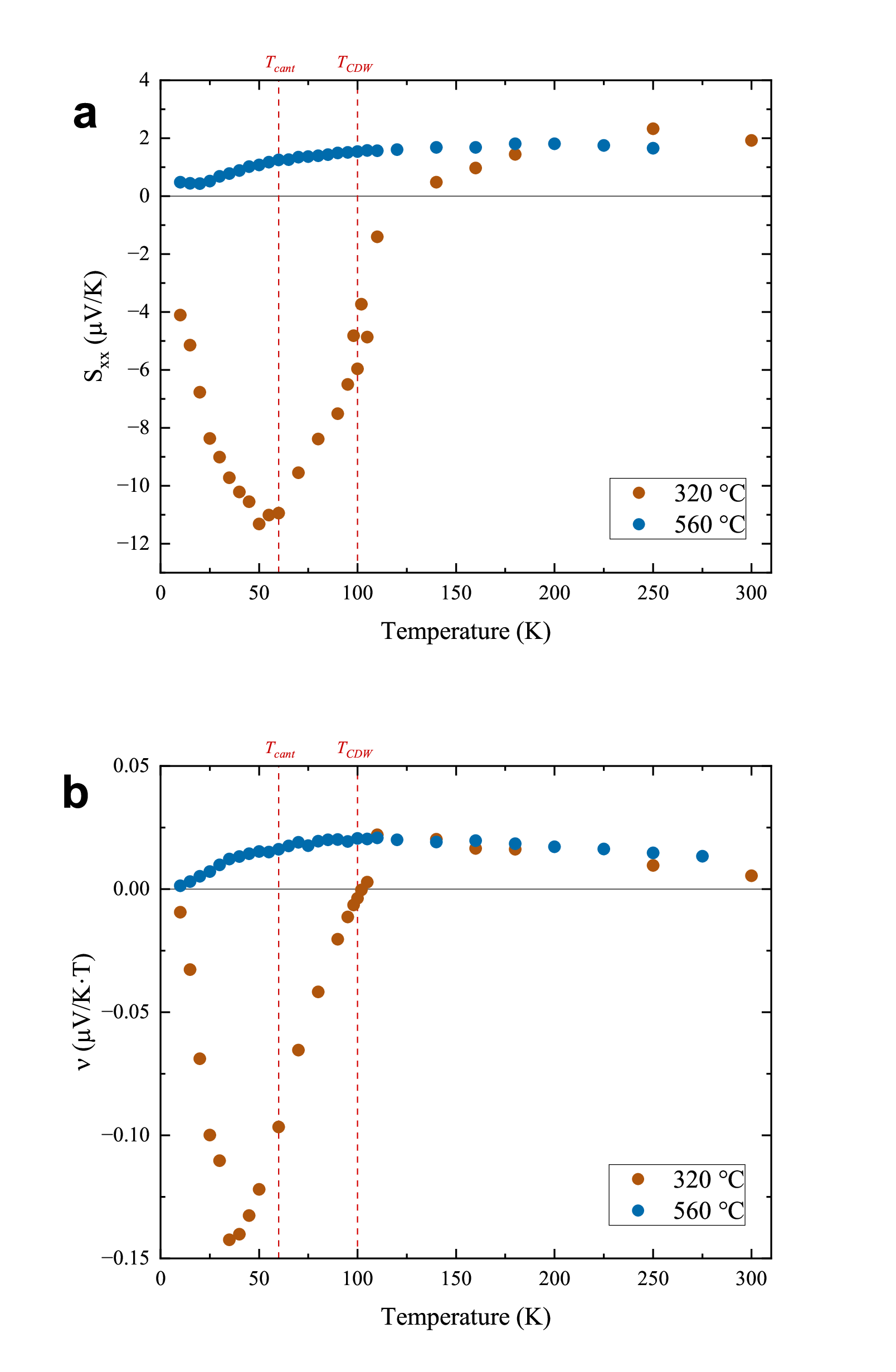}
		\caption{(a) Seebeck coefficient $S_{xx}$ and (b) Nernst coefficient $\nu$ for the 320 $\degree$C annealed sample (orange) and the 560 $\degree$C annealed sample (blue) in the magnetic field up to 14 T. The heat current is applied along the ab-plane, and the Seebeck data points were obtained from the zero-field $\Delta$V value extracted from field sweeps at fixed temperatures, normalized by the sample geometry and the temperature difference across the sample.}
		\label{Fig3}
	\end{center}
\end{figure}

Further analysis of these trends can be observed in Fig. \ref{Fig3}a, which displays the zero-field theropower $S_{xx}$ as a function of temperature for each annealing condition, and Fig. \ref{Fig3}b, which shows $\nu$ for each annealing condition as a function of temperature. Both figures show remarkably similar features in both $S_{xx}$ and $\nu$, with the 560 $\degree$C annealed sample remaining exclusively positive, and the 320 $\degree$C annealed sample changing sign close to $T = T_{CDW}$. Both of these samples share hole-like charge carriers at room temperature, which we define here as a positive Seebeck signal. 

Furthermore, a rapid slope change appears to occur in the 320 $\degree$C annealed sample due to the charge density wave transition, and $S_{xx}$ exhibits a kink feature at $T_{CDW}$, which we observe close to 105 K. This sign change is clearly absent in the sample annealed at 560 $\degree$C. In the CDW-containing sample, this trend continues for the temperature range $T_{CDW}$ $ < T < 300$ K that was measured. At $T < T_{CDW}$, the  sample extablishes much sizable negative (electron-like) thermpower signal, confirming that the dominant charge carriers are electrons for the CDW state. Holes become the dominant quasiparticle charge carriers around 130 K, when $S_{xx}$ is positive in Fig. \ref{Fig3}a.

\begin{figure}[!htb]
	\begin{center}
		\includegraphics[ width= \columnwidth]{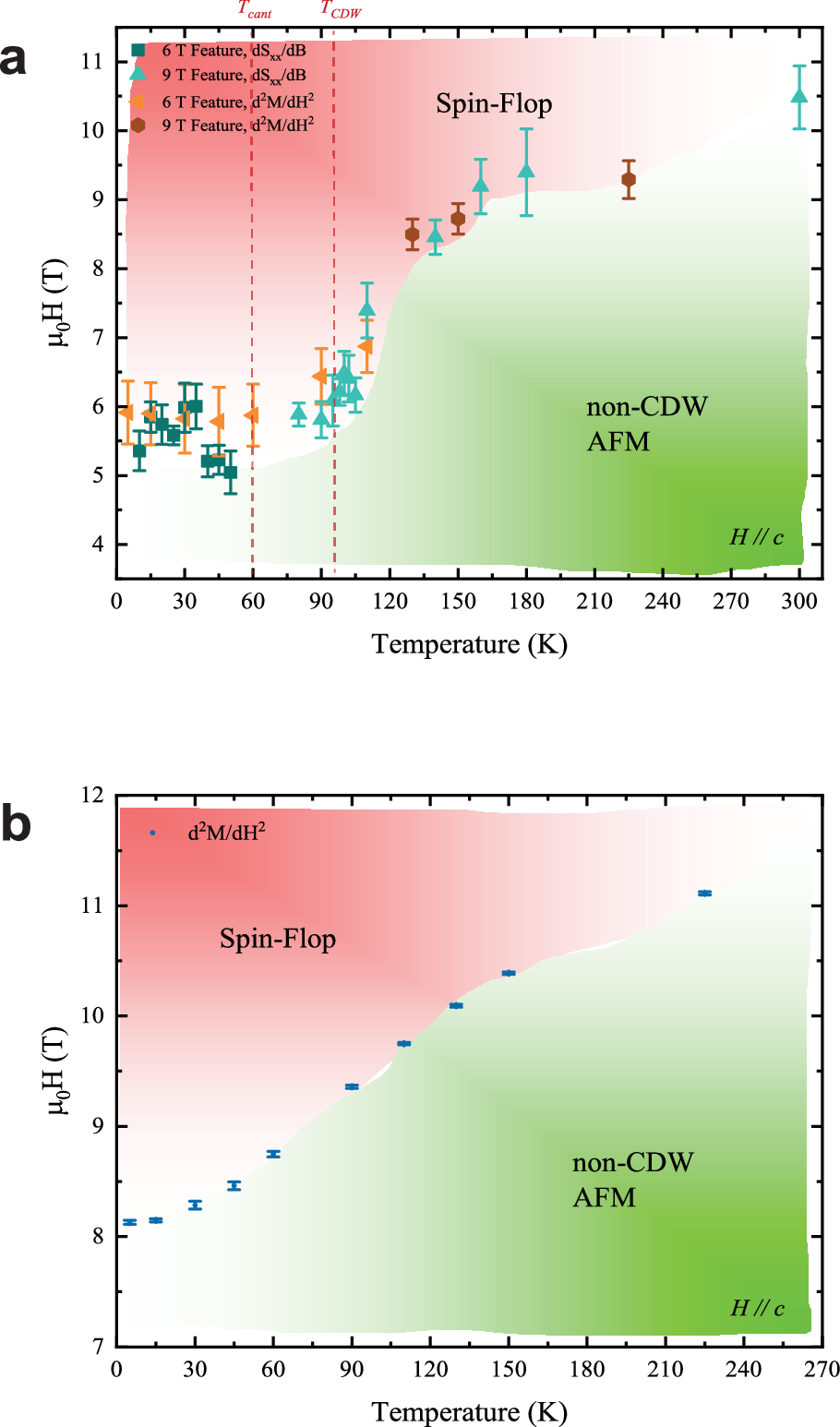}
		\caption{ Thge phase diagram of $\mu_0H$ vs. $T$ for (a) the 320 $\degree$C annealed sample and (b) the 560 $\degree$C annealed sample, using the locations of various magnetic features to define the phase transition. The boundaries of the canted AFM phase and CDW-ordered phase are deliniated here as $T_{cant}$ and $T_{CDW}$ respectively.}
		\label{Fig4}
	\end{center}
\end{figure}

Similarly, the Nernst coefficient $\nu$ established a similar trend in $T$. For $T < T_{CDW}$, the sample exhibiting the CDW shows a much more sizable Nernst coefficient in the opposite sign than that after the CDW transition. The Nernst coefficient $\nu$-$T$ is plotted in Fig. \ref{Fig3}b. Notably, $\nu$ for the 320 $\degree$C annealed sample is significantly larger than that of the 560 $\degree$C annealed sample. This signal has a peak around 35-40 K, and changes sign around 105 K. We believe this sign change is further indicative of the CDW present in the 320 $\degree$C annealed sample and is again not present in $\nu$ for the 560 $\degree$C sample here. The absolute magnitude of the Nernst coefficient in the 320 $\degree$C annealed sample is almost 22x larger than the 560 $\degree$C sample at its largest magnitude as well. We also note that the 320 $\degree$C annealed sample has much stronger incommensurate magnetic peaks associated with spin density wave order below around 60 K that are considerably suppressed for the 560 $\degree$C annealed sample \cite{Klemm2024}. 

These findings are further examined in the phase diagrams seen in Fig. \ref{Fig4}a and b. Regarding the 320 $\degree$C annealed sample seen in Fig. \ref{Fig4}a, there are two dM/dH peaks observed at 110 K, but only one peak for the temperatures both above and below 110 K (see the supplement for the magnetization $M-H$ data). This feature in addition to the Nernst sign change around 105 K both provide evidence for a spin-flop transition region between 105 and 110 K, near the CDW feature, consistent with previous work \cite{Klemm2024}. Previous works have also asserted the presence of incomensurate magnetic Bragg peaks around 60 K \cite{Teng2022, Teng2023,Wu2024}, which is also supported by our measurements and labeled here as $T_{cant}$.
It is interesting to point out that recent photoemission work has also identified changes in the electronic structure at this characteristic temperature scale in the 320 $\degree$C annealed sample, while absent in the 560 $\degree$C annealed sample~\cite{Oh2025}, consistent with the trends of $S_{xx}$ and $\nu$ shown here.

In conclusion, in the Kagome metal FeGe, this thermoelectric effect study demonstrates a dramatic change in electronic structure accompanying the CDW transition. Compared to the crystal without the CDW transition, the crystal with the CDW is revealed to have a significantly enhanced Nernst effect and a dramatic sign change of the dominating carrier in the thermopower effect. This result suggests that the disorder created by the annealing conditions in
Ge-1 site may prevent the CDW formation \cite{Klemm2024}. Thermoelectric effects are also shown to detect the spin-flop transition.

{\it Acknowledgement.} The work at the University of Michigan is supported by the National Science Foundation under Award No. DMR-2004288 and DMR-2317618 (transport measurements), by the Department of Energy under Award No. DE-SC0020184 (magnetization measurements) to Kuan-Wen Chen, Dechen Zhang, Guoxin Zheng, Aaron Chan, Yuan Zhu, Kaila Jenkins, and Lu Li. The single crystal synthesis and characterization at Rice are supported by US NSF DMR-2401084 and the Robert A. Welch Foundation under Grant No. C-1839 (P.D.).

\bibliographystyle{ieeetr}

\end{document}